\documentclass[prb,twocolumn,showpacs]{revtex4}
\usepackage{amsmath,amsfonts}
\usepackage{graphicx}
\begin{document}
\title{The self-assembly of DNA Holliday junctions studied with a minimal model} 
\author{Thomas E.~Ouldridge}
\affiliation{Rudolf Peierls Centre for Theoretical Physics, 1 Keble
        Road, Oxford, UK OX1 3NP}
\author{Iain G.~Johnston}
\affiliation{Rudolf Peierls Centre for Theoretical Physics, 1 Keble
        Road, Oxford, UK OX1 3NP}
\author{Ard A.~Louis}
\affiliation{Rudolf Peierls Centre for Theoretical Physics, 1 Keble
        Road, Oxford, UK OX1 3NP}
\author{Jonathan P.~K.~Doye}
\affiliation{Physical and Theoretical Chemistry Laboratory, 
  Department of Chemistry, University of Oxford, 
  South Parks Road, Oxford OX1 3QZ, United Kingdom}
\date{\today}

\begin{abstract}
In this paper, we explore the feasibility of using coarse-grained models to simulate
the self-assembly of DNA nanostructures. We introduce a simple model of DNA
where each nucleotide is represented by two interaction 
sites corresponding to the sugar-phosphate backbone and the base. 
Using this model, we are able to simulate the self-assembly of both 
DNA duplexes and Holliday junctions from single-stranded DNA.
We find that assembly is most successful in the temperature window 
below the melting temperatures of the target structure and 
above the melting temperature  of misbonded aggregates. 
Furthermore, in the case of the Holliday junction, we show how a hierarchical 
assembly mechanism reduces the possibility of becoming trapped in misbonded 
configurations.
The model is also able to reproduce the relative melting 
temperatures of different structures accurately, 
and allows strand displacement to occur.
\end{abstract}
\maketitle

\newcommand{\av}[1]{\left\langle#1\right\rangle}

\section{Introduction}
The ability to design nanostructures which accurately self-assemble from 
simple units is central to the goal of engineering objects and machines 
on the nanoscale. 
Without self-assembly, structures must be laboriously constructed in a step 
by step fashion.  Double-stranded DNA (dsDNA) has the ideal properties for 
a nanoscale building block,\cite{Seeman2003,Pitchiaya2006} 
with structural length scales determined by the 
separation of base pairs, the helical pitch and its persistence length 
(approximately 0.33\,nm, 3.4\,nm (Ref.\ \onlinecite{Saenger1984}) and 
50\,nm,\cite{Hagerman1988} respectively). 
Over these distances, dsDNA acts as an almost rigid 
rod and so it is capable of forming well-defined three dimensional structures.

It is the selectivity of base pairing between single strands, however, that makes DNA ideal for controlled self-assembly. 
By designing sections of different strands to be complementary, a certain configuration of a system of oligonucleotides can be specified as the global minimum of the energy landscape. 
In this way the target structure (usually consisting of branched double helices) can be `programmed' into the sequences. 
This approach was initially demonstrated for a four-armed junction by the Seeman
group in 1983.\cite{Kallenbach83}
Such junctions and more rigid double crossover motifs\cite{Fu93}
can then be used to create two-dimensional lattices.\cite{Winfree98,Malo2005}
Yan \it et al. \rm  \cite{Yan2003} have also constructed ribbons and two dimensional lattices from larger four-armed structures, each arm consisting of a junction of four strands.  
Furthermore, using Rothemund's DNA ``origami'' approach an almost arbitrary 
variety of two-dimensional shapes can be created.\cite{Rothemund06}

Progress in forming three-dimensional DNA nanostructures was initially much 
slower. The Seeman goup managed to synthesize a DNA cube\cite{Chen91} and 
a truncated octahedron,\cite{Zhang94} but only after a long series of steps and 
with a low final yield.
More recently, approaches have been developed that allow polyhedral cages, 
such as tetrahedra,\cite{Goodman2005} trigonal bipyramids,\cite{Erben07} 
octahedra,\cite{Shih04,Anderson08} dodecahedra and truncated icosahedra,
\cite{He2008} to been obtained in high yields simply by cooling 
solutions of appropriately designed oligonucleotides from high temperature. 
Additional structures have also been produced using pre-assembled modular building blocks incorporating other organic molecules.\cite{Aldaye2007,Zimmermann08} 

In designing strand sequences, it is important to minimize the stability of competing structures with respect to the stability of the target configuration. 
In addition, if systems can be designed to follow certain routes through configuration space---for example, by the hierarchical assembly of simple motifs \cite{Pistol2006}---the target can potentially be reached more efficiently. 
A standard approach to hierarchical assembly, such as that described by 
He {\it et al.},\cite{He2008}
involves choosing sequences so that bonds between different pairs of oligonucleotides become stable at different temperatures. 
This allows certain motifs to form in isolation at high temperatures before bonding to each other as the solution is cooled. 
An alternative, elegant system for programming assembly pathways has been proposed by Yin \it et al. \rm \cite{Yin2008}, which relies on the metastability of single stranded loop structures and the possibility of catalyzing their interactions using other oligonucleotides.

Given these recent experimental advances in creating DNA nanostructures, it would be useful to have
theoretical models that allow further insights into the self-assembly process.
In particular, a successful model would be able to provide information on the 
formation pathways and free energy landscape associated with the self-assembly, 
and as such would be of use to experimentalists wishing to consider increasingly more complex designs. 
Atomistic simulations of DNA would offer potentially the most spatially-detailed descriptions of the self-assembly.  
However, they are computationally very expensive, and are generally restricted to time scales that are 
too short to study self-assembly.\cite{Cheatham2004}

Statistical approaches such as that of Poland and Scheraga \cite{Poland1970} and the nearest-neighbour model \cite{Everaers2007} use simple expressions for the free energy of helix and random coil states to obtain equilibrium results for the bonding of two strands. 
Whilst the parameters in these models can be tuned to give very accurate correspondence with experimental data,\cite{SantaLucia1998,SantaLucia2004} they give no information on the dynamics and formation pathways and hence are only useful for ensuring that the target structure has significantly lower free energy than competing configurations. 
Furthermore, any description purely based on secondary structure (i.e.\ which bases are paired) is inherently incapable of accounting for topological effects such as linking of looped structures.\cite{Bois2005} 

Coarse-grained or minimal models offer a compromise  between detail and computational simplicity, and are well  suited to the study of hybridization of oligonucleotides.  The aim of these models is to be capable of describing  both the thermodynamic and kinetic behaviour of systems, a vital feature if kinetic metastability is inherent in assembly pathways.\cite{Yin2008} In developing such minimal models the approach is usually to retain just those physical features of the system that are essential to the behaviour that is of interest. 

Dauxois, Peyrard and Bishop models,\cite{Dauxois1993} and modified versions such as that proposed by Buyukdagli, Sanrey, and Joyeux,\cite{Buyukdagli2005} 
constitute the simplest class of dynamical models. 
Although these are dynamic models in the sense that the energy is a function of the separation between each base, the nucleotides are constrained to move in one dimension. This lack of conformational freedom means that these models are 
incapable of capturing the nuances of the self-assembly from single-stranded DNA (ssDNA).

Recently, models have been proposed which capture the helicity of dsDNA using two \cite{Drukker2001} or three \cite{Knotts2007}  interaction sites per nucleotide. 
These models, however, are optimized for studying deviations from the ideal double-stranded state, and so have not been used to examine self-assembly.
Although they have been used to study the thermal denaturation of dsDNA, 
it is essential for our purposes to be able to simulate the assembly of a 
structure from ssDNA as it is this process that will reveal the kinetic traps 
and free energy landscape associated with the formation of a particular 
DNA nanostructure.  

Simpler,  linear models, also with two interaction sites per nucleotide, have been used to investigate duplex hybridization,\cite{Araque2006} hairpin formation \cite{Sales-Pardo2005} and gelation of colloids functionalized with oligonucleotides.\cite{Starr2006} 
These models all use two interaction sites to represent one nucleotide, with backbone sites linked to each other to represent the sugar-phosphate chain, and interaction sites which represent the bases. This work investigates the possibility of extending the use of such coarse-grained models to study the self-assembly of nanostructures that involve multiple strands forming branched duplexes. We hypothesize that the self-assembly properties of DNA are dominated by the fact that ssDNA is a semi-flexible polymer with selective attractive interactions. 
We introduce an extremely simple model, similar to that of Starr and Sciortino,
to test this hypothesis. \cite{Starr2006}  
This simplicity enables us to explore the thermodynamics and kinetics of self-assembly in the model in great depth, and hence examine the feasibility of simulating nanostructure formation through minimal models.

We first describe the model in Section \ref{methods}, then examine its success in reproducing the general features of hybridization in Section \ref{Duplex Formation}. 
Next in Section \ref{Holliday Junction}, we apply it to the formation of a Holliday junction, a simple nanostructure consisting of a four-armed cross.\cite{Malo2005}

\begin{figure}
\begin{center}
\includegraphics[width=8.4cm]{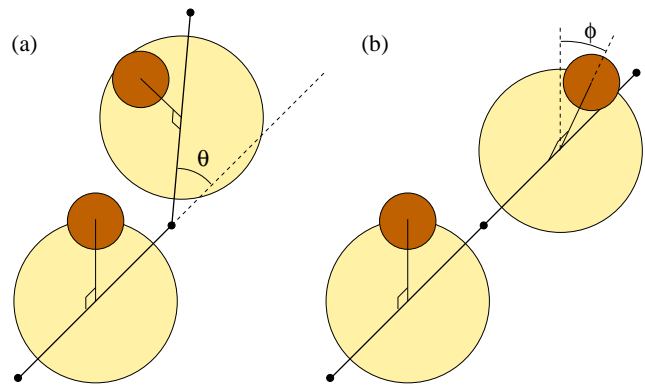}
\end{center}
\caption{(Colour online) A schematic representation of the model. 
The thick lines represent the rigid backbone monomer units and the large circles the repulsive Lennard-Jones interactions at their centres.
The smaller, darker circles represent the bases. 
The panels illustrate the definitions of (a) the bending angle between two units ($\theta$), and (b) the torsional angle ($\phi$) which is found after the monomers have been rotated to lie parallel.}
\label{model picture}
\end{figure}

\section{Methods}
\label{methods}
\subsection{Model}
\label{model}

We introduce an off-lattice model inspired by that which Starr and Sciortino 
used to study the gelation of four-armed DNA dendrimers.\cite{Starr2006} 
As our aim is to reproduce the basic physics with as simple a 
model as possible, we neglect contributions to the interactions
due to base stacking, and the charge and asymmetry of the phosphate backbone. 
We do not attempt to include the detailed geometrical structure 
of DNA, but instead
represent the oligonucleotides as a chain of monomer units,
each corresponding to one nucleotide (Fig.\ \ref{model picture}). 
A monomer consists of a rod (chosen to be rigid for simplicity) of length $l$ 
with a repulsive backbone interaction site at the centre of the rod.
In addition, each unit has a bonding interaction site (or base) at a distance 
of $0.3\,l$ from the backbone site 
(perpendicular to the rod). Each monomer is also assigned a base type 
(A,G,C,T) to model the selective nature of bonding. In this model we only 
consider bonds between the complementary pairs A-T and G-C.

We do not explicitly include any solvent molecules in our simulations, but instead use effective potentials
to describe the interactions between the DNA.
Sites interact through shifted-force 
Lennard-Jones (LJ) potentials, where, as well as truncating and shifting the 
potential, an extra term is included to ensure the force goes smoothly to zero 
at the cutoff $r_c$. 
For $r < r_{c}$
 \begin{equation} 
V_{\rm sf}(r)=V_{\rm LJ }(r) - V_{\rm LJ}(r_{c}) - 
              (r-r_{c})\left.\frac{dV_{\rm{LJ}}}{dr}\right|_{r=r_c}
 \label{potential} 
 \end{equation} 
where
 \begin{equation} 
 V_{\rm LJ}(r) = 4\epsilon \left [\left({\sigma \over r} \right) ^{12} - \left( {\sigma \over r} \right) ^{6} \right], \label{LJ} 
\end{equation} 
and $V_{\rm sf}(r) = 0$ for $r\ge r_{c}$. 
Backbone sites (except adjacent units on the same strand) interact through 
Eq.\ (\ref{potential}) with $\sigma = l$ and $r_{c} = 2^{1/6}\sigma$. 
This purely repulsive interaction models the steric repulsion between strands. 
Bonding sites (again excluding adjacent units on a strand) interact via 
Eq.\ (\ref{potential}) with $\sigma = 0.35\,l$ and  $r_{c} = 2.5\,\sigma$ for complementary bases (to allow for attraction) and $r_{\rm c} = 2^{1/6}\sigma$ for all other pairings. 
The depth of the resulting potential well between complementary bases, 
$\epsilon^{\rm eff}_{\rm base}$, is $0.396\epsilon$. In what follows we will measure
the temperature in terms of a reduced temperature, 
$T^*=k_B T/\epsilon^{\rm eff}_{\rm base}$. 
The above choice of parameters ensures that the attractive interaction between complementary bases is largely shielded by backbone repulsion. Monomers therefore bond selectively and can only bond strongly to one other monomer at a given instant. 
These are the key features of Watson-Crick base pairing that make DNA so useful for self-assembly.

The model also includes potentials between consecutive monomers associated with bending and twisting the strand: 
\begin{equation} 
V_{\rm bend} = \begin{cases}
                k_1 (1-\cos(\theta))& 
                        \text{if $\theta < {3\pi\over4}$},\\ 
		\infty& \text{otherwise}
			\end{cases}
\label{bend} 
\end{equation}
and 
\begin{equation}
V_{\rm twist} = k_2 (1-\cos(\phi)).
\label{twist}
\end{equation}
We define $\theta$ as the angle between the vectors along adjacent monomer rods. As previously mentioned, consecutive backbone sites do not interact via LJ potentials. 
Instead, a hard cutoff is introduced in Eq.\ (\ref{bend}) to reflect the fact that an oligonucleotide cannot double back on itself. 
$\phi$ is taken as the angle between adjacent backbone to bonding site vectors after the monomers have been rotated to lie parallel (Fig.\ \ref{model picture}). 
For simplicity, we choose the torsional potential to have a minimum at $\phi=0$. Thus, neither ssDNA or dsDNA
will be helical in our model.

$k_1$ is chosen to be $0.1\,\epsilon$ to give a persistence length, $l_{\rm ps}=3.149\,l$ at a reduced temperature of $T^*=0.09677$ for ssDNA. 
We obtain this result by simulating a single strand 70 bases in length, and using the definition:\cite{Cifra2004}
\begin{equation}
l_{\rm{ps}} = \frac{\langle {\bf{L} \cdot \bf{l_{\rm 1}}}\rangle} {\langle l_{\rm 1} \rangle},
\end{equation} 
where $\bf{L}$ is the end to end vector of the strand, $\bf{l_{\rm 1}}$ is the vector associated with the first monomer and $\langle \rangle$ indicates a thermal average. 
Taking $l$ to be 6.3{\AA}, $T^*=0.09677$ is mapped to $24^\circ$C and so the model is consistent with experimental data for ssDNA in 0.445M NaCl solution.\cite{Murphy2004,lengthscale} $k_2$ is chosen to be $0.4\,\epsilon$.

In neglecting the geometrical structure of a double helix we do not accurately represent certain types of bonding. 
`Bulged' bonding occurs when consecutive bases in one of the strands attach to non-consecutive bases in the other strand. 
`Internal loops' consist of stretches of non-complementary bases (either symmetric or asymmetric in the number of bases involved in each strand). 
`Hairpins' result when a single strand doubles back and bonds to itself. The details of these motifs are complicated but an empirical description of their thermodynamic properties is given in Ref.\ \onlinecite{SantaLucia2004}. 
Importantly, they are generally penalized due to the disruption of the geometry of DNA in a way which is not well reproduced by our model. 
In the case of the short strands we consider, these motifs will only play a small role as the strands are not specifically intended to have stable structures of these forms. 
In fact, as the base sequences we use were designed to form the Holliday junction, the possibility of forming these motifs at relevant temperatures was deliberately avoided.\cite{MaloThesis}

For simplicity we therefore include only two alterations to the model. Firstly, we define `kinked states' as those for which the number of unpaired bases between two bonding pairs on either side of a duplex is not equal (including asymmetric loops and bulges). We impose an infinite energy penalty on the formation of these kinked states if the total number of intermediate bases is less than six. Secondly, we treat complementary units within six bases of each other on the same strand as non-complementary, but allow all other hairpins without penalty. 

It should be also noted that this model neglects the directional asymmetry of the sugar-phosphate backbone. Therefore,
parallel as well as anti-parallel bonding is possible in our model, whereas parallel bonding does 
not occur in experiment.    

\subsection{Monte Carlo Simulation}
\label{Monte Carlo Simulation}

In a fully-atomistic model of DNA the natural way to simulate its dynamics would be to use molecular dynamics. However, the best way to simulate the dynamics
in a coarse-grained model is an important, but not fully resolved, question, and
one that will depend on the nature of the model. Clearly, for the current model
standard molecular dynamics is inappropriate as it will lead to ballistic 
motion of the strands between collisions because of the absence of explicit 
solvent particles, whereas DNA in solution undergoes diffusive Brownian motion. 
An alternative approach is to use Metropolis Monte Carlo (MC) 
algorithm\cite{Metropolis1953,Frenkel2001} where the moves 
are restricted to be local, as it has been argued that this can provide a 
reasonable approximation to the dynamics.\cite{Kikuchi91,Berthier07,Tiana07} 
This is the approach that we use here to simulate the dynamics of 
self-assembly of DNA duplexes and Holliday junctions. 
In particular, the local MC moves that we use are translation and 
rotation of whole strands and bending of a strand about a particular monomer,
thus ensuring that the strands undergo an approximation to diffusive Brownian 
motion in the simulations.
Therefore we expect the MC simulations, which are all initiated with free single strands, to mimic the real self-assembly processes in our model. It is 
important to note that this will include, as well as successful assembly into the target structure, kinetic trapping in non-equilibrium
configurations and that when the latter occurs this reflects the inefficiency of
the self-assembly under those conditions.

Although a true measure of time is impossible in Monte Carlo simulations, an approximate time scale for diffusion-limited processes can be found by comparing the diffusive properties of objects to experiment. 
By measuring the diffusion of isolated strands, and assuming diffusion coefficients comparable to those of double strands and hairpin loops of similar length,\cite{Lapham1997} we conclude that one step per strand corresponds to a time scale of approximately 2\,ps. 
Thus, our model allows our systems to be studied on millisecond time scales. 

At the end of the above MC simulations, our systems will not necessarily have reached equilibrium, 
both because the energy barriers to escape from misbonded configurations can be difficult to 
overcome at low temperature and the low rate of association at higher 
temperatures.
Therefore, as a comparison we also compute
the equilibrium thermodynamic properties of our systems using umbrella 
sampling.\cite{Torrie1977,Frenkel2001} 
Formally, we can write the thermal average of a function $B(\bf{r}^N)$ in the canonical ensemble as:
\begin{equation}
\langle{B}\rangle = \frac
   {\int{\frac{B}{W(Q)} \left[{W(Q) \exp(-V/\rm{k_B}\it T)}\right]} d\bf{r}^{N}} 
   {\int{\frac{1}{W(Q)}\left[{W(Q) \exp(-V/\rm{k_B} \it T)}\right]} d\bf{r}^{N}} 
\label{US avg}
\end{equation}
where $Q=Q(\bf{r}^N)$ is an order parameter or reaction coordinate and $V=V(\bf{r}^N)$ is the potential energy. 
We are free to choose $W(Q)$, and by taking the term in square brackets as the weighting of states and keeping statistics for $B/W$ and $1/W$ at each step we can find $\langle{B}\rangle$. 
In standard Metropolis MC, $W=1$, but by choosing $W(Q)$ in such a way that those states 
with intermediate values of $Q$ are visited more frequently, the effective free energy barrier between 
(meta)stable states can be lowered allowing the system to pass easily between the free energy minima, and equilibrium to be reached.

To ensure that each value of $Q$ is equally likely to be sampled in an 
umbrella sampling simulation, one would choose $W(Q)=\exp(\beta A(Q))$,
where $A(Q)$ is the free energy as a function of the order parameter. 
To achieve this, however, would require knowledge of $A(Q)$. 
Instead, there are standard methods to 
construct $W(Q)$ iteratively, but for the current examples it was possible to
construct $W(Q)$ manually, because of the relative simplicity of the free
energy profiles.

To a first approximation the interaction between fully bonded structures is negligible. 
Therefore, in the umbrella sampling simulations we consider systems containing the minimum number of strands 
required to form a given object (two for a duplex and four for a Holliday junction). We then use the relative weight of bound and free states to extrapolate the expected fractional concentrations for larger systems.\cite{Ouldridgeunpub}
The natural choice for the order parameter $Q$ is the number of correct bonds, where 
two monomers are defined to be bonded if their energy of interaction is negative.

\section{Results}
\subsection{Duplex Formation}
\label{Duplex Formation}

\begin{figure}
\includegraphics[width=6.2cm,angle=-90]{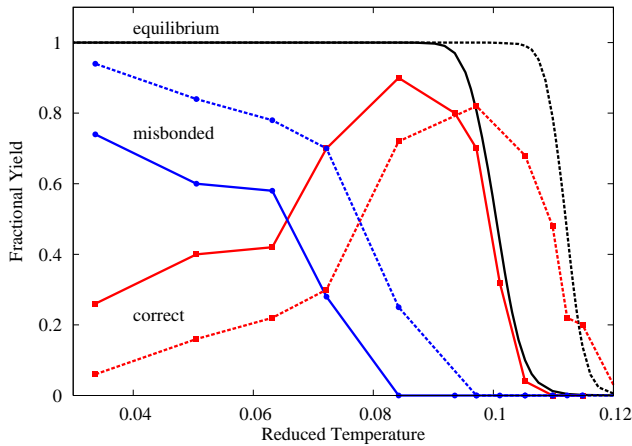}
\caption{(Colour online) Yields of correctly-formed duplexes and misbonded configurations at the end 
of our MC simulations (lines with data points, as labelled) compared to the equilibrium probability of 
the strands adopting the correct structure as obtained by umbrella sampling.
The solid and dashed lines represent results for strands with 7 and 13 monomers, respectively.
The MC results are averages over ten runs of length $3\times10^8$ steps per strand with ten strands in 
the simulation cell. }
\label{7 & 13 mers}
 \end{figure}

We test the model by analysing the duplex bonding of two different complementary strands. 
We simulate systems of ten oligonucleotides, initially not bonded, in a periodic cell with a concentration of $5.49\times10^{-5}$ molecules\,$l^{-3}$
(or $3.65\times10^{-4}$\,M). 
We separately consider strands consisting of 7 and 13 monomers, which correspond to two of the arms of the Holliday junction studied experimentally by Malo {\it et al.}\cite{MaloThesis,Malo2005} and which we consider in Section \ref{Holliday Junction}:
\begin{equation}
\text{7 bases} \begin{cases}
					\text{G-A-G-T-T-A-G}\\
					\text{C-T-A-A-C-T-C}
					\end{cases}
\label{7 bases}					
\end{equation}
\begin{equation}					
\text{13 bases} \begin{cases}
					\text{G-C-G-A-T-G-A-G-C-A-G-G-A}\\
					\text{T-C-C-T-G-C-T-C-A-T-C-G-C}
					\end{cases}
\label{13 bases}
\end{equation}
where we have listed strands in the 5'--3' sense for consistency with the literature. 
The yields of correctly-bonded and misbonded structures at the end of the simulations are depicted in Figure 
\ref{7 & 13 mers} as a function of temperature.
We also display the predicted equilibrium fraction of correctly bonded strands for both systems obtained using umbrella sampling. 
For convenience, we define a correctly bonded structure to have more than 70\% of the bonds of the complete duplex and no bonds to other strands. Any other structure is recorded as `misbonded'.

\begin{figure}
\includegraphics[width=8.4cm]{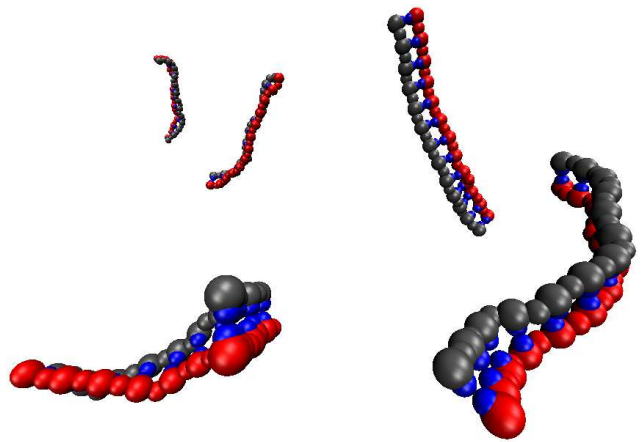}
\caption{(Colour online) Snapshot of a fully-assembled configuration in a MC simulation of
ten 13-base strands  at $T^*=0.0971$.
In this image the colour of the backbone indicates the type of strand: red for G-C-G-A-T-G-A-G-C-A-G-G-A and grey for its complement.
Backbone sites are indicated by the large spheres, and bases by the small, blue spheres.}
\label{5 13mers}
\end{figure}

Figure \ref{7 & 13 mers} shows a maximum in the yield as a function of temperature. 
Such behaviour is typical of self-assembling systems 
\cite{Brooks06,Hagan2006,Wilber2007,Rapaport08,Whitelam08} and reflects the 
thermodynamic and dynamic constraints on the self-assembly process. 
Firstly, the yield is zero at high temperature where only ssDNA is stable, and rises just below the expected 
equilibrium value as the temperature is decreased, 
the deviation arising due to the large number of steps required to reach equilibrium. 
At low temperatures, the yield falls away due to the presence of kinetic traps which are now stable with respect to isolated strands, as evidenced by the rise in `misbonded structures' in Figure \ref{7 & 13 mers}. 
Thus, there is a non-monotonic dependency of yield on temperature and an optimum region for successful assembly, which corresponds to the region where only the desired structure is stable against thermal fluctuations. 
Figure \ref{5 13mers} is a snapshot from near the end of a simulation in this regime. 
It should be noted that neglecting helicity has the effect of increasing the flexibility of dsDNA in the direction 
perpendicular to the plane of bonding 
(a helix cannot bend in any direction without disturbing its internal structure whereas a `ladder' can). 

\begin{figure}
\includegraphics[width=8.4cm]{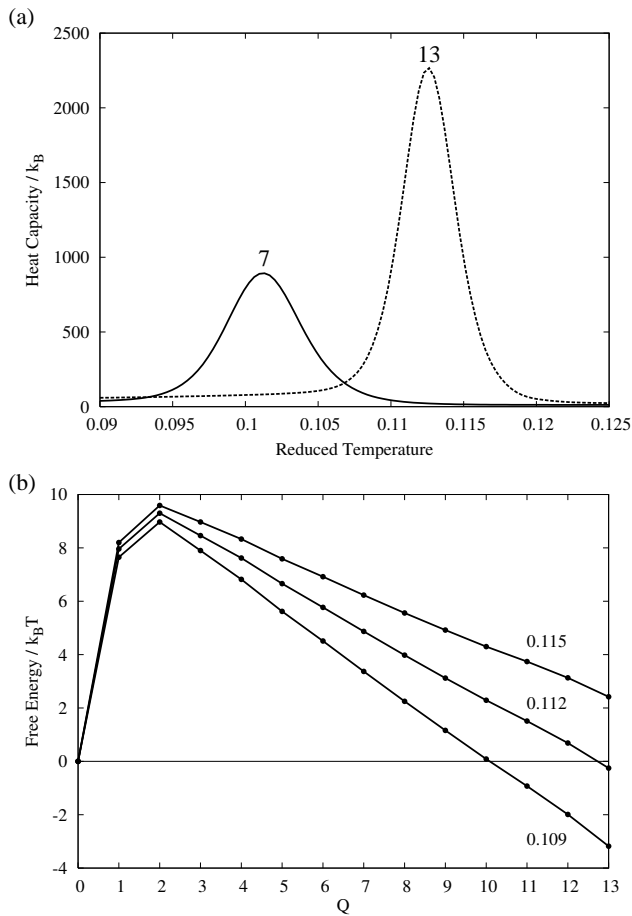}
\caption{Thermodynamics for the formation of a single duplex. 
(a) Heat capacity curves for 7- and 13-base systems, as labelled.
(b) Free energy profile associated with the formation of a 13 base pair duplex at different temperatures 
(as labelled), where $Q$ represents the number of correctly formed bonds in the duplex.}
\label{thermo_duplex}
\end{figure}

The heat capacity obtained from umbrella sampling of pair formation is shown in Figure \ref{thermo_duplex}(a). 
The heat capacity peaks indicate a transition from single strands to a duplex. 
As the formation of duplexes is essentially 
a chemical equilibrium between monomers and clusters of a definite size 
(in this case two)
the width of the peaks will remain finite as the number of strands is increased. 
The transition does, however, become increasingly narrow as the DNA strands become longer, as is evident from comparing the 
heat capacity peaks for the 7-mer and 13-mers.

Figures \ref{thermo_duplex}(b) shows the free energy profile, $F(Q)$, for the 
formation of a duplex. 
The initial peak at low $Q$ is accounted for by the entropic cost of bringing two strands together. 
Once bonds are formed, however, adding extra bonds costs much less entropy whilst providing a significant decrease in energy, explaining the monotonic decrease in $F(Q)$ beyond $Q=2$. 
The rise between $Q=1$ and $Q=2$ is partly due to the fact that in order to form two bonds between strands the relative orientation of strands must be specified whereas this is not true for $Q=1$: 
hence there is an additional entropy penalty to the formation of the second bond. 
In addition, there exist structures with only one correct bond that are 
stabilized by additional incorrect bonds and these misbonded configurations 
also contribute to $F(1)$. 
The constant gradient above $Q=2$ indicates that the energetic gain and entropic cost of forming an extra bond are approximately constant at a given temperature, which is consistent with the assumptions underlying nearest-neighbour models of DNA melting.\cite{Everaers2007}

\begin{figure}
\includegraphics[width=6.2cm,angle=-90]{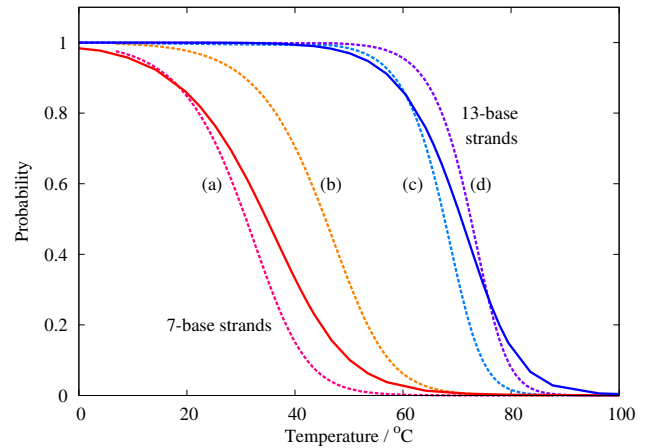}
\caption{(Colour online) The bulk equilibrium probability of strands being in 
a correct duplex extrapolated from our umbrella sampling simulations (solid lines) compared to the predictions of an empirical two-state model (dashed lines).
Results are presented for strands with 7 and 13 bases, as labelled.
For the two-state model, as well as the results for the sequences
in Eqs.\ (\ref{7 bases}) and (\ref{13 bases}) (lines (a) and (d)),
sequences corresponding to the other two arms of the Holliday junction
(Figure \ref{HJ schematic}) are considered.
Only one line is shown for the umbrella sampling results, because A-T and G-C
have the same binding energy in our model.
Temperatures in our model are converted to  ${\rm^o}$C using the same mapping given in Section II A.}
\label{2 state}
\end{figure} 

 In Figure \ref{2 state}, we compare our melting curves to those predicted by a simple two-state model,\cite{Everaers2007} using the same mapping of the reduced temperature as in Section \ref{model}. In the two-state model the molar concentrations of product (AB) and reactants (A,B) are given by the equilibrium relation: 
\begin{equation}
\frac{\left[{\rm AB }\right]}{\left[{\rm A} \right] \left[{\rm B}\right]} = \exp{\left({\frac{-\Delta H_0 + T \Delta S_0}{\rm{k_B}\it T}}\right)},
\label{two state model}
\end{equation}
where $\Delta H_0$ and $\Delta S_0$ are assumed to be constants which depend only on the strand sequences and the salt concentration (we take $\rm{[Na^+}] = 0.445\rm{M}$ as in Section \ref{model}). 
We use the enthalpy and entropy changes of duplex formation calculated by ``HyTher",\cite{Hyther} a program that estimates these values using the ``unified oligonucleotide nearest neighbour parameters".\cite{SantaLucia1998, Peyret1999} 
The authors claim that the thermodynamic parameters predicted by HyTher give the melting temperature $T_{\rm m}$ (the temperature at which the fraction of bonded strands is 1/2)
of a duplex to within a standard error of $\pm2.2^{\rm{o}}\rm{C}$.\cite{SantaLucia1998}
Figure \ref{2 state} shows that our system reflects the melting temperatures predicted by the two-state model with reasonable accuracy, excepting sequence dependent effects which are not included in our model, because the interaction energies 
between A-T and C-G complementary base pairs have for simplicity been taken to be the same. The widths of transitions are seen to be of the same order, but slightly larger for our model. 
This feature, which is typical of coarse-grained models,\cite{Knotts2007} indicates that the degree of entropy loss on hybridization is too small in our model, 
and is due to a failure to accurately incorporate all degrees of freedom which become frozen on hybridization. 
However, the agreement is sufficiently good that the basic features of physical DNA assembly should be reproducible.

\begin{figure*}
\begin{center}
\includegraphics[width=15cm]{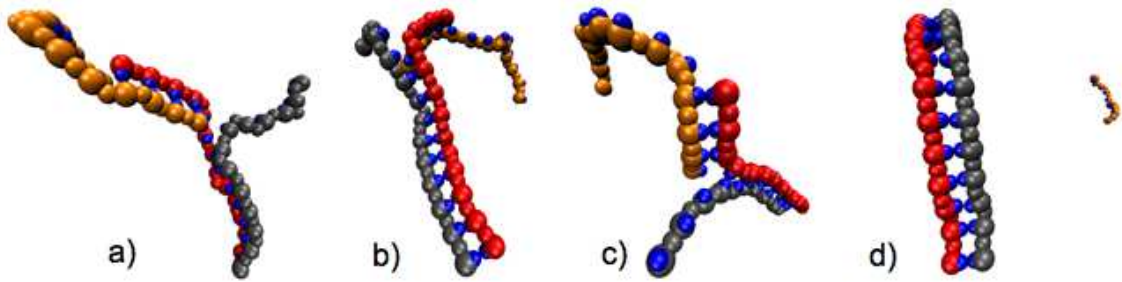}
\end{center}
\caption{(Colour online) Snapshots illustrating four stages in the process of displacement. 
(a) A third strand binds to a misbonded pair. 
(b) The third strand is prevented from forming a complete duplex by the misbond. 
(c) Thermal fluctuations cause bonds in the misbonded structure to break and be replaced by the correct duplex. 
(d) The misbonded strand is displaced and the correct duplex is formed.}
\label{displacement}
\end{figure*}

A further satisfying feature of the model is that `displacement' was observed on several occasions. This process, during which a misbonded pair of strands is broken up by a third strand, is illustrated in Figure \ref{displacement}. The third strand is able to bond to the pair, as some bases are free in the misbonded structure. 
Thermal fluctuations allow the new strand to bond to sites previously involved in misbonding, in a process known as `branch migration'. Eventually one of the misbonded strands is completely displaced, leaving a correct duplex and an isolated single strand. This behaviour is observed in real DNA systems, and is the driving mechanism of some nanomachines \cite{Yurke2000} and DNA catalyzed reactions.\cite{Yin2008, Zhang2007}

\begin{figure}
\begin{center}
\includegraphics[width=8.4cm]{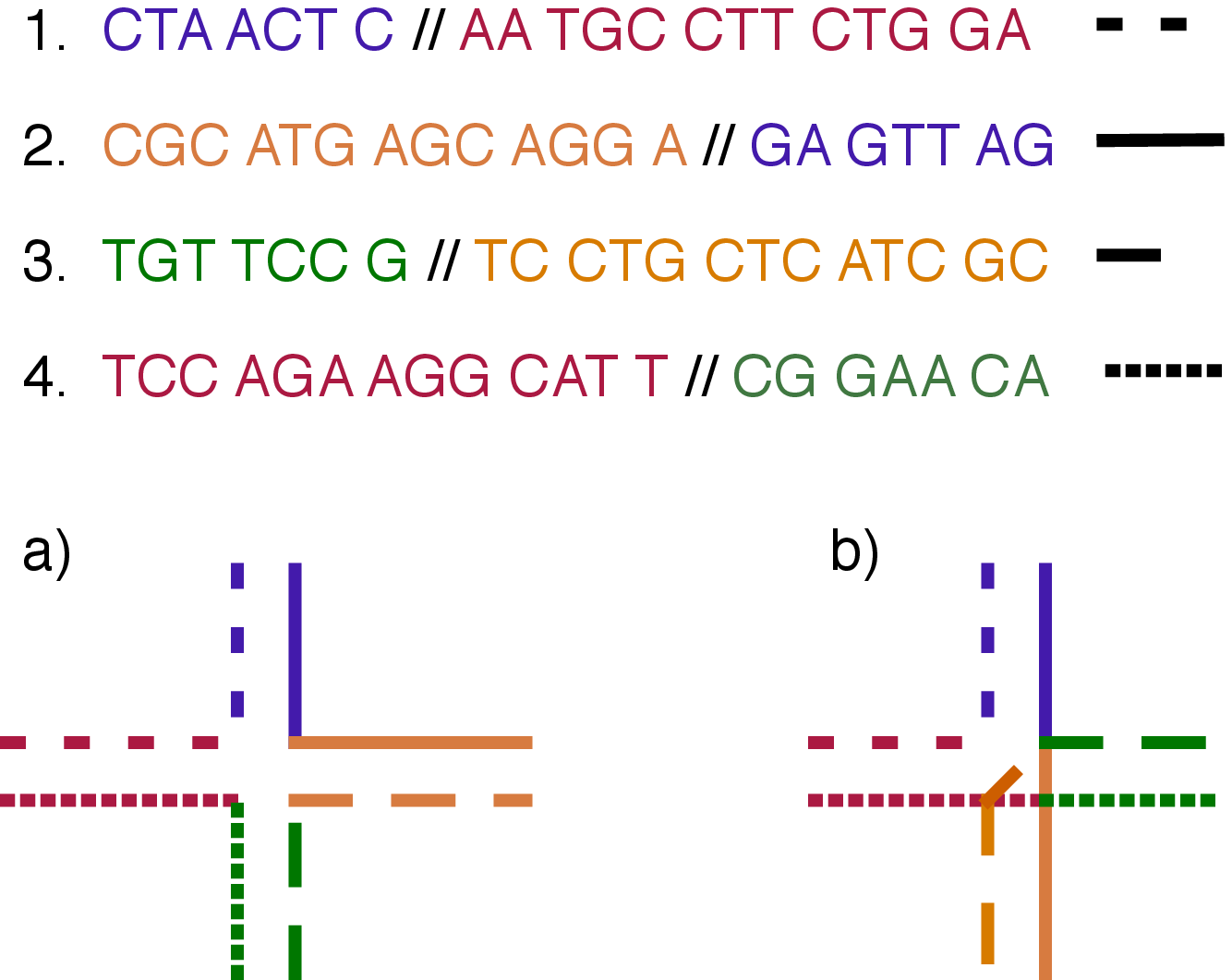}
\end{center}
\caption{(Colour online) A schematic diagram showing the sequences of the strands used in our Holliday junction simulations, and the alternative bound states that are possible: (a) the square planar configuration and (b) the $\chi$-stacked form.}
\label{HJ schematic}
\end{figure}

\subsection{Holliday Junction}
\label{Holliday Junction}
Encouraged by the above results,
we next apply the model to the formation of a Holliday junction. 
Holliday junctions consist of four single strands which bind to form a four-armed cross. 
In our case we consider a Holliday junction with two long arms (13 bases long) and two short arms (7 bases long). 
We use the experimental base ordering of Malo {\it et al.}\cite{MaloThesis} with the `sticky ends' removed. 
(These sticky ends consist of six unpaired bases on the end of arms and their purpose is to allow the Holliday junctions to bond together to form a lattice). 
The sequences of the four DNA strands and schematic diagrams of the 
possible junctions that they can form are shown in Figure \ref{HJ schematic}.

\begin{figure}
\includegraphics[width=6.0cm,angle=-90]{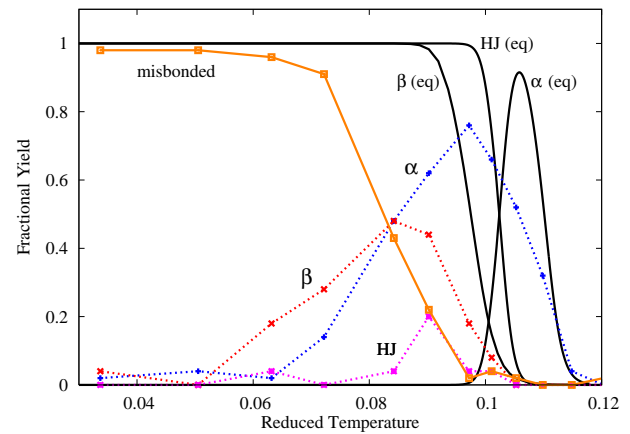}
\caption{(Colour online) 
A comparison of the kinetics and thermodynamics for a system of 20 strands that 
can potentially form five Holliday junctions, where the MC simulations
are initiated from a purely single-stranded configuration. 
The MC results (lines with data points) are the final yield of Holliday junctions, and the fraction of strands involved in a correctly-formed long ($\alpha$) or 
short ($\beta$) arm, or in misbonding, as labelled.
The results are averages over five runs of length $10^{9}$ steps per strand.
For comparison, the equilibrium probabilities of being in a $\alpha$-bonded dimer
and a Holliday junction are also plotted, along with the equilibrium probability of 
being in a $\beta$-bonded dimer if the longer arms are not allowed to hybridize.
}
\label{20mers}
\end{figure}

Initially we studied a system of 20 strands (five of each type) that has the potential to form five separate junctions. 
We use a concentration of $1.56 \times 10^{-5}$ molecules\,$l^{-3}$
(which corresponds to $1.04 \times 10^{-4}$M). 
The results are displayed in Figure \ref{20mers}. 

The results are as expected for the bonding of the longer arms (which we now describe as `$\alpha$-bonding'). 
The yield again displays the characteristic non-monotonic dependence on temperature. 
We obtain very few complete junctions, however, which is due to two effects. 
Firstly, each simulation is performed at constant temperature, which means the hierarchical route to assembly is less favoured than when the system is cooled, as in the experiments.\cite{MaloThesis,Malo2005}
When the system is gradually cooled, Figure \ref{20mers} suggests that at around $T_{\rm m}(\alpha) =0.111$ we would expect to find a region in which only $\alpha$-bonded dimers were stable with respect to ssDNA. 
If the cooling was sufficiently slow on the timescale of bonding, all strands would form $\alpha$-structures at around $T_{\rm m}(\alpha)$. 
At lower temperatures, when the Holliday junction becomes stable with respect to the $\alpha$-bonded dimers, many competing minima would then be inaccessible to the system as they would require the disassociation of stable $\alpha$-bonded pairs. 
The free energy landscape of two $\alpha$-structures forming a Holliday junction is consequentially much simpler than that of four single strands forming a junction at a given temperature.  
Therefore, one expects the yield for self-assembly at constant temperature to be 
lower than when the system is cooled, because there is only a relatively narrow 
temperature window between where the Holliday junction becomes stable and misbonded
configurations start to appear. Indeed, the shorter arms are only marginally more 
stable than some competing minima, as evidenced by 
the rise in misbonded structures in Fig.\ \ref{20mers} at temperature just below 
where $\beta$-bonded structures first appear.

\begin{figure}
\begin{center}
\includegraphics[width=8.4cm]{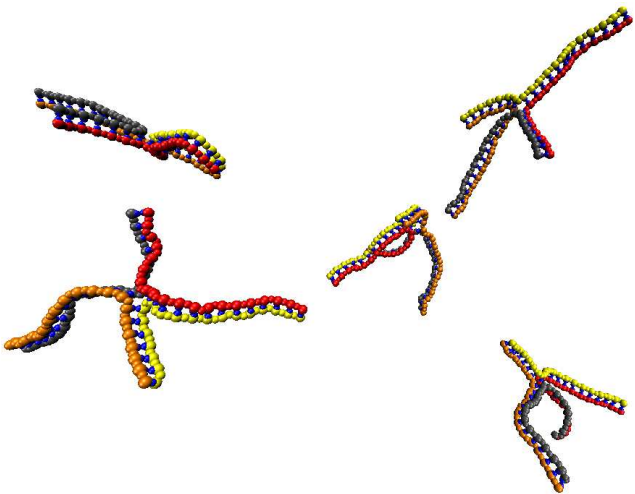}
\end{center}
\caption{(Colour online) Snapshot showing five Holliday junctions formed at $T=0.0842$ after 
$5.67 \times 10^8$ MC steps per strand. Again, the backbone colour indicates strand type (1: red, 2: grey,
 3: orange, 4: yellow) where numbers refer to Figure \ref{HJ schematic}}
\label{5 HJs}
\end{figure}

Secondly, even in the temperature range where Holliday junction formation is 
not hindered by the formation of misbonded configurations, the yield is low
because the Metropolis MC algorithm artificially reduces the diffusion of bound 
pairs, and hence the likelihood that two pairs of $\alpha$-bonded strands come together to form a junction is also reduced.
This is because the acceptance probability of trial moves for bonded strands is much lower than for isolated strands \cite{Luijten2006}, due to the energy penalty associated with trying to move a bound pair apart. 

\begin{figure}
\includegraphics[width=6.0cm,angle=-90]{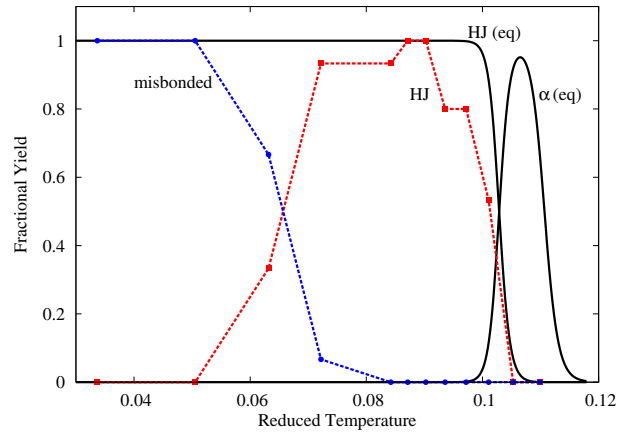}
\caption{(Colour online) The yields of Holliday junctions (HJ) and misbonded configurations 
for MC simulations, where the initial configuration was a pair of 
$\alpha$-bonded dimers. For comparison, 
the equilibrium probabilities of being in a $\alpha$-bonded dimer 
and a Holliday junction are also plotted.
The results are averages over five runs of length 
$7.5 \times 10^8$ steps per strand.
}
\label{HJ thermo}
\end{figure}

Interestingly, examination of the equilibrium lines in Figure \ref{20mers} shows that the Holliday junctions are actually stable at a higher temperature than the individual shorter arms. 
This is because the total loss of entropy when two $\alpha$-bonded dimers bind together is considerably less than that for two short arms in isolation (as fewer translational degrees of freedom are lost), whereas the energy change is comparable. 
Thus, there is a small temperature window at $T^*\approx 0.1$ where hierarchical assembly can occur at constant temperature as the short arms are only stable once $\alpha$-bonding has taken place. 
However, due to the deficiencies in the MC simulations mentioned above, the yield
of Holliday junctions in this region is practically zero. Instead, the maximum
yield of Holliday junctions occurs at lower temperatures where non-hierarchical
pathways that proceed by the addition of single strands become feasible.

The above simulations were only able to successfully model the first stage of 
the Holliday junction assembly, namely the formation of $\alpha$-bonded dimers. 
To probe the second stage of assembly, we must first make two modifications to 
our simulation approach to overcome the two deficiencies mentioned above. 
Firstly, we study systems initially consisting of pairs of $\alpha$-bonded strands, which we assume have successfully formed at some higher temperature---this is 
reasonable given the results of our earlier simulations.
Secondly, we also include simple local cluster moves in addition to those which move only one strand, i.e.\ translations, rotations and bending of pairs of $\alpha$-bonded strands. 
With these changes incorporated, we simulate the same system for $7.5 \times 10^8$ steps per strand at a range of temperatures below $T_{\rm m}(\alpha)$. 
It should be noted that due to a change in the size of typical moves, one move per strand now corresponds to approximately 10ps.

We find that Holliday junctions form over a wide range of intermediate temperatures, whilst kinetic traps at low temperature lead to incomplete bonding and consequently to the possibility of forming large clusters. A typical result from the high-yield regime is shown in Figure \ref{5 HJs}. 
As fully-bonded Holliday junctions are essentially inert, 
it is reasonable to analyse their assembly behaviour by considering only one 
junction.  The smaller system size has the effect of increasing the assembly rate, because the strands have less distance to diffuse, but does not affect the basic assembly mechanism. 
We therefore simulated systems consisting of two $\alpha$-bonded pairs with the same concentration as above. 

We also introduced some modifications to the the umbrella sampling scheme in order
to more efficiently compute the thermodynamics of the second-stage of 
Holliday junction formation. 
As well as cluster moves, we also introduced a `tethering' component in the 
weighting function $W(Q)$. 
We introduce a length $r_{\rm min}$ that corresponds to the shortest distance between any pair of backbone sites on different strands. 
We then split $Q=0$ into two regions: we weight those states with  $r_{\rm min} < 3l$ with $W=1$ but for  $r_{\rm min} \geq 3l$ we use $W=0.1$. 
This enables us to increase the rate of transitions between $Q=0$ and 1, and reduces the time spent 
simply simulating the diffusion of $\alpha$-bonded dimers waiting for a collision to occur.

The MC results are plotted in Figure \ref{HJ thermo} along with 
the equilibrium results obtained from umbrella sampling.
With the cluster moves in place, we now see a high yield of Holliday junctions
and a broad maximum in the yield as a function of temperature.
The hierarchical pathway has the effect suggested earlier. 
Namely, the temperature window over which correct formation can occur is 
vastly increased, as the most significant competing minima are inaccessible because their formation would require dissociation of the $\alpha$-bonded pairs. 

The model is therefore consistent with the experimentally observed hierarchical assembly of Holliday junctions as the system is cooled.\cite{Malo2005, MaloThesis} 
It should be noted, however, that the junctions in our model usually form in the `square planar' as opposed to the `$\chi$-stacked' shape (Figure \ref{HJ schematic}) that is observed under normal experimental conditions. 
The preference for one structure is a subtle consequence of the concentration of cations and the precise helical geometry of DNA.\cite{Ortiz-Lombardia1999} 
This level of detail is not included in our coarse-grained model, so it is not surprising that it cannot reproduce the preference for $\chi$-stacked structures.
Moreover, it is relatively easy to see why in our model, which forms 
`ladders' rather than helices, a planar geometry is preferred for the junction.

\begin{figure}
\includegraphics[width=8.4cm]{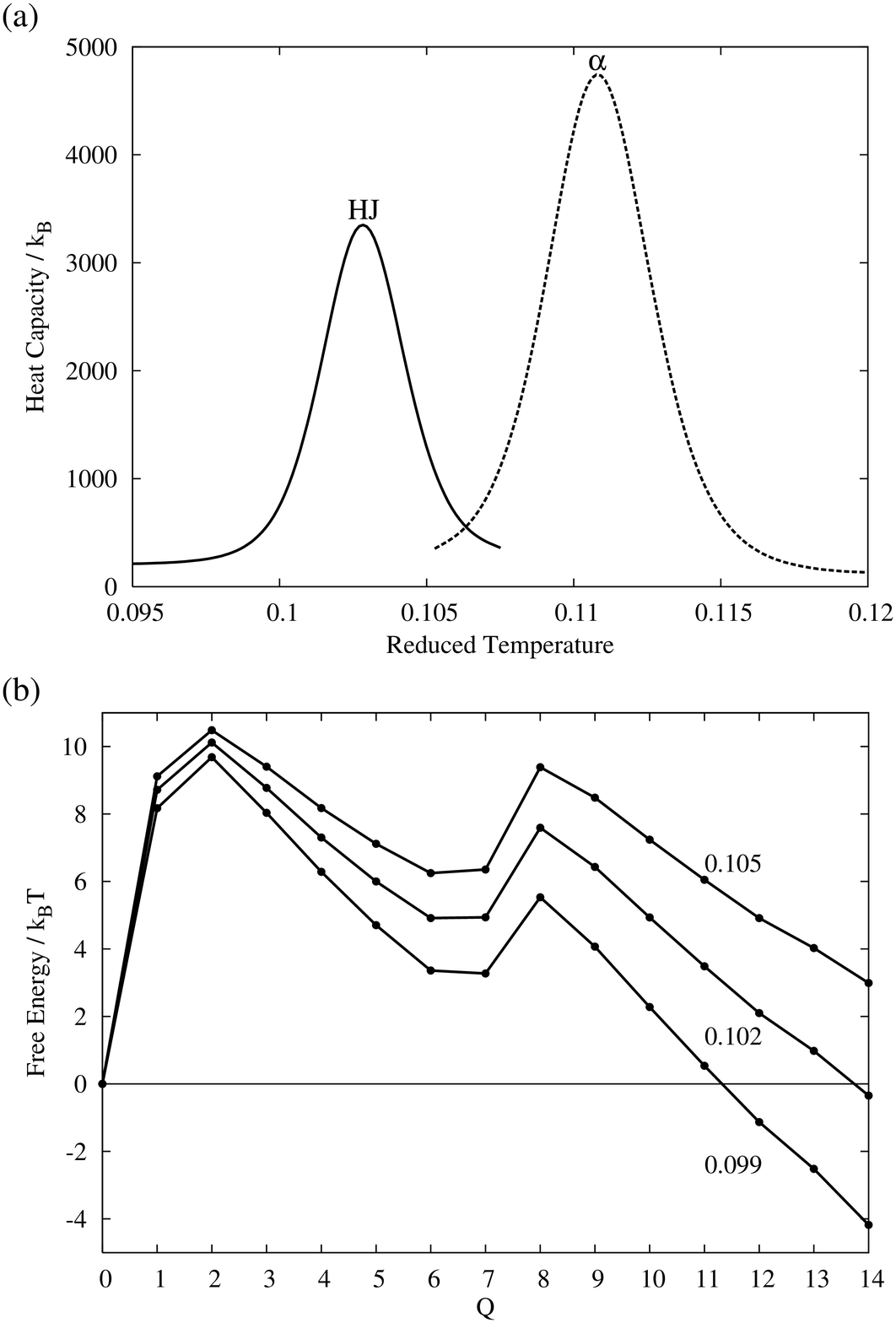}
\caption{Thermodynamics for the formation of a single Holliday junction.
(a) Heat capacity curves for two $\alpha$-bonded dimers forming a Holliday junction (HJ) 
and four strands forming two $\alpha$-bonded dimers, as labelled.
(b) Free energy profiles for the formation of a Holliday junction from two $\alpha$-bonded dimers 
at different temperatures, as labelled. 
$Q$ represents the total number of correct bonds in the short arms of the junction.
}
\label{thermo_HJ}
\end{figure}

Some of the equilibrium thermodynamic properties associated with the formation
of a Holliday junction are shown in Figure \ref{thermo_HJ}. 
In particular, Fig.\ \ref{thermo_HJ}(b) shows the free energy profile for the 
formation of a Holliday junction from two $\alpha$-bonded pairs. 
The initial peak and subsequent drop is very similar to that for the duplexes and can be accounted for in the same way. 
However, the formation of the two arms is not like the zipping up of a 
14-base duplex, because there is much more relative freedom of movement for 
the bases on either side of the $\alpha$-bonded sections in the dimers
than for consecutive bases on single-stranded DNA.
Thus, there is a rise between $M=7$ and $M=8$ that is a result of the entropy 
penalty of bringing together the two ends to make the second short arm. 
We note that the penalty is much smaller than the initial cost of bringing the two $\alpha$-bonded pairs together, and as a result, 
the value of $T_{\rm{m}}$ for the junction is higher than for the short arms 
in isolation, as noted earlier.

An interesting feature of Figure \ref{thermo_HJ}(b) is the plateau between $Q=6$ and $Q=7$. 
In general, when two $\alpha$-bonded pairs meet to form one short arm, there is an entropic penalty associated with the excluded volume that the remaining bases in the $\alpha$-structures represent to each other. 
This excluded volume is a large fraction of the total available space if one complete short arm is formed, so that there are no free monomers between the short arm and the $\alpha$-bonded sections. As a consequence, there is not the usual free energy benefit from forming the final bond in the short arm (the one closest to the centre of the Holliday junction), as the excluded volume penalty is large and those states that are allowed involve distortion of the backbones and bonds near the centre of the Holliday junction. 
Although the details of this free energy penalty and the other features in 
Fig.\ \ref{thermo_HJ}(b) will depend on the exact geometry of the system, 
we expect the calculated free energy profile to be representative of that 
for real DNA.

It is possible to extend the two-state model discussed in Section \ref{Duplex Formation} to the formation of a Holliday junction by considering the concentrations of all four isolated strands, the two $\alpha$-bonded intermediates and the junction itself. 
We assume Eq.\ (\ref{two state model}) holds for every possible transition, use the same thermodynamic parameters as before and apply conservation of total strand number. To estimate $\Delta H_0$ and $\Delta S_0$  associated with the formation of a Holliday junction, we construct a single strand by linking the ends of the oligonucleotides together with four non-bonding bases. The thermodynamic parameters associated with the folding of this structure are predicted by ``UNAFold''.\cite{Markham2008} 
The correction for the fact that our strands are not connected by loops is discussed by Zuker.\cite{Zuker2003} 
This leaves five simultaneous equations (assuming perfect stochiometry) which can be solved numerically.

\begin{figure}
\includegraphics[width=6.2cm,angle=-90]{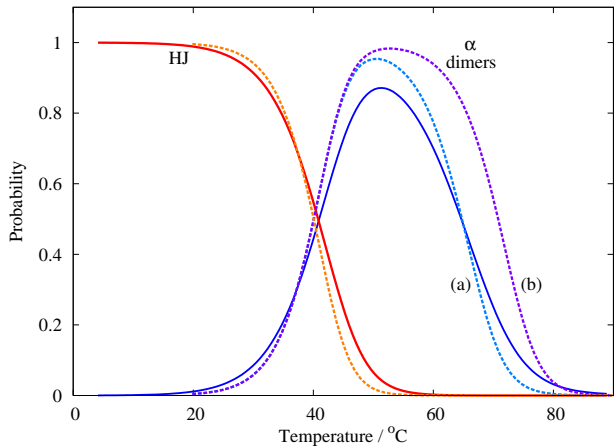}
\caption{(Colour online) Bulk equilibrium probability of strands being in a Holliday junction (HJ) or
an $\alpha$-bonded dimer computed by umbrella sampling (solid lines) and by 
the extended two-state model (dashed lines), where lines (a) and (b) represent the two possible $\alpha$-bonded dimers.
}
\label{etsm HJ}
\end{figure}

Figure \ref{etsm HJ} compares this extended two-state model (ETSM) with the bulk thermodynamics predicted by umbrella sampling (using the same temperature scaling as before). 
ETSM predictions for both stages of Holliday junction formation agree well with our results, which again supports our hypothesis that much of the physics of self-assembly can be reproduced by a simple coarse-grained model. The extra width of the transitions in our model occurs for the same reasons as mentioned in Section \ref{Duplex Formation} when discussing Fig.\ \ref{2 state}. 

\begin{figure}
\includegraphics[width=6.1cm,angle=-90]{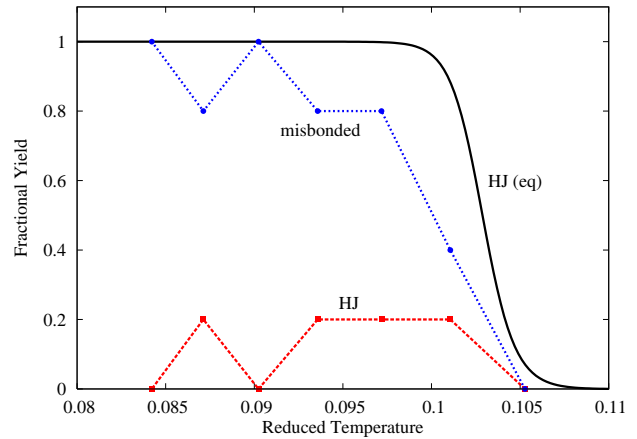}
\caption{(Colour online) Simulation results for the badly-designed Holliday 
junction of Section \ref{negative}, where the initial configuration was a pair 
of $\alpha$-bonded dimers. The lines with data points give the yield of 
correctly-formed junctions and misbonded configurations, as labelled.
For comparison the solid line gives the equilibrium probability that the 
well-designed sequences of Section \ref{Holliday Junction} adopt a 
Holliday junction.}
\label{thermo neg}
\end{figure}

\begin{figure}
\begin{center}
\includegraphics[width=7.4cm]{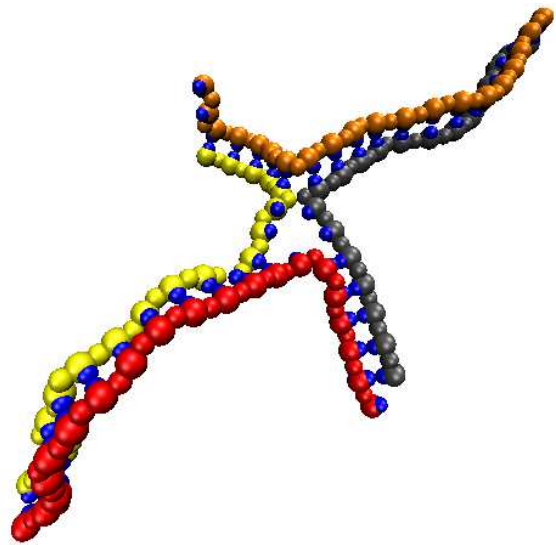}
\end{center}
\caption{(Colour online) Example of a competing minimum for a badly designed Holliday junction. The snapshot is 
taken from a simulation at $T^*=0.0936$. 
}
\label{misbond neg}
\end{figure}

\subsection{Negative Design}
\label{negative}
The hierarchical pathway for the formation of the Holliday junction is one 
aspect of the sequence design that aids the formation of the correct structure.
The experimental base ordering of the Holliday junction, however, was also chosen to minimize the number of competing structures---a typical example of `negative design'.\cite{Doye04b} 
We illustrate the importance of such negative design by considering a badly-designed junction, where the complementary seven-base sections consist of just one base type each. 
We simulate a system of two $\alpha$-bonded pairs with the short arms so modified 
under the same conditions as for Fig.\ \ref{HJ thermo} and 
the results are shown in Figure \ref{thermo neg}. 
Although there is some probability of forming the correct junction, the simulations
are dominated by misbonded junctions, such as the one depicted in Figure \ref{misbond neg}.
Although these competing structures are energetically less stable than the target
junction because of the presence of unpaired bases at the `dangling' ends,
they are readily accessible, because the likelihood that the first bonds formed
between two $\alpha$-bonded pairs are in the same registry as the target structure  
is low. The yield of the correct Holliday junctions will then depend upon how
readily the system is able escape from these malformed junctions. Clearly,
this process is slow on the time scales of the current simulations, and is 
also likely to hinder the location of the target structure in experiment.

\section{Discussion}
In this paper we have introduced a simple coarse-grained model of DNA in order to test the feasibility of modeling the self-assembly of DNA nanostructure by 
Monte Carlo simulations. Any such model involves a trade-off between detail and 
computational simplicity, and here we deliberately chose to keep the model as simple
as possible in order to give us the best chance of being able to probe the time
scales relevant to self-assembly. The model involves just two interaction sites
per nucleotide. 

The results from our model are very encouraging. Firstly, we have shown that
using our model it is feasible to model the self-assembly of both DNA duplexes
and a Holliday junction. The latter represents, to the best of our knowledge, the
first example of the simulation of the self-assembly of a DNA structure beyond a 
duplex. Secondly, the model succeeds in reproducing many of the known thermodynamic
and dynamic features of this self-assembly. 
For example, the equilibrium melting curves agree well with those predicted by 
the nearest-neighbour two-state model,\cite{SantaLucia1998} which is known
to predict melting temperatures very accurately. 
The model is also able to capture important dynamical 
phenomena such as displacement.

Thirdly, by analysing the thermodynamic and dynamic constraints on assembly, we 
have been able to gain some important physical insights into the nature of DNA
self-assembly and how to control it. For example, the optimal conditions for 
self-assembly are in the temperature range just below the melting temperature of the
the target structure, where this structure is the only one stable with respect to 
the precursors, be they ssDNA or some intermediate in a hierarchical assembly 
pathway.  At lower temperatures, misbonded configurations can be formed that act 
as kinetic traps and reduce the assembly yield.
Similar trade-offs between the thermodynamic driving force and 
kinetic accessibility have been previously seen in a variety of self-assembling
systems,\cite{Brooks06,Hagan2006,Wilber2007,Rapaport08,Whitelam08} 
and also give rise to a maximum in the yield near to and below
the temperature at which the target structure becomes stable.

We have also seen how hierarchical self-assembly through cooling can be a 
particularly useful strategy to aid self-assembly, because the formation of stable
intermediates at higher temperatures simplifies the free energy landscape for the
assembly of the next stage in the hierarchy by reducing the number of misbonded 
configurations available to the system. This simplification of the energy
landscape is likely to be a general feature of hierarchical self-assembly.

Thus, our results have confirmed the utility of using coarse-grained DNA models to study the self-assembly of DNA nanostructures, and supported our hypothesis that much of the physics can be explained by describing DNA as a semi-flexible polymer with selective attractive interactions. The model's success in forming junctions in reasonable computational time suggests that it will be possible to develop further models that have an increased level of detail, but which can still access the time scales relevant to self-assembly.

The model has also highlighted some features which it would be advantageous to include in such models. For example, greater accuracy in the details of oligonucleotide geometry, particularly the helicity of dsDNA, would allow features such as the characteristically long persistence length of hybridized strands to be reproduced and give the appropriate degree of rigidity to simulated nanostructures. Such improvement might also allow more complicated motifs to be accounted for, such as the preference for $\chi$-stacked Holliday junctions that the current model could not reproduce. 

It should be noted that if one is to introduce helicity in a physically
reasonable way it should also allow for ssDNA to undergo a stacking transition 
to a helical form. 
This transition may play a significant role in the thermodynamics and kinetics of self-assembly.\cite{Holbrook1999}
Previously proposed coarse-grained DNA models that incorporate helicity have not been designed to accurately reproduce this feature. 
Incorporating extra degrees of freedom which are relevant to the stacking transition, such as the rotation of the base with respect to the sugar-base bond, may also help to increase the entropy change on hybridization and hence make the transition narrower as required.

The approximation to diffusive dynamics provided by the local move Metropolis Monte Carlo algorithm could also be improved. 
Currently the `local' moves involve displacing, rotating or bending entire strands or pairs of strands---these effectively constitute cluster moves of groups of strongly bound nucleotides, and result in slow relaxation and translation times within bound structures. 
More realistic dynamics may be achievable by considering trial moves of individual nucleotides, and incorporating cluster moves in a more systematic fashion, such as in the `virtual move' MC algorithm proposed by Whitelam and Geissler.\cite{Whitelam2007,Whitelam08} 

One potential issue with any coarse-graining is how it preserves the different
time scales in a system.
In Section \ref{Monte Carlo Simulation} we assigned an approximate mapping 
between the number of Monte Carlo steps and physical time based upon comparison 
of diffusion coefficients. There are, however, other important time scales in the 
system, such as the time scale for the internal dynamics of an isolated strand and 
the time scale over which the `zipping-up' of two strands occurs after a bond has been formed. 
Comparisons of experimental diffusion coefficients \cite{Lapham1997} and 
melting and bubble formation from molecular dynamics 
simulations \cite{Drukker2001,Knotts2007} suggest a large separation in time scale 
between diffusion-limited processes and those that rely on the dynamics of 
individual nucleotides. 
Encouragingly, we observe a similar time scale separation in our model:
zipping-up and thermal relaxation of isolated strands occur over times scales 
shorter than $10^5$ steps per strand, whereas association typically required on the 
order of $10^7$ to $10^8$ steps per strand near the melting temperature 
(corresponding to tens or hundreds of microseconds). Furthermore, we would
argue that it is this time scale separation, and not the precise ratios of the
relevant rate constants, that it is important to reproduce in self-assembly 
simulations.

We should also note that the mapping of the diffusion constants between the model
and experiment will not necessarily ensure that the rate of association is 
accurate in our model, because although the frequency of collisions in our model 
should be correct, 
there is also the contribution
to the association rate from the probability that a collision will lead to 
successful association. That we can reproduce the thermodynamics of the DNA melting transitions implies that the rates of association and disassociation have the right
ratio, but not that they necessarily have the correct absolute value. 
For example, it is conceivable that helicity (both in dsDNA and possibly in ssDNA),
which is not included in the current model, will influence the likelihood that a 
collision is successful.

\begin{acknowledgments}
The authors are grateful for financial support from the EPSRC 
and the Royal Society. We also wish to acknowledge helpful discussions
with Jonathan Malo, John Santalucia Jr and Michael Zuker.
\end{acknowledgments}


\begin{thebibliography}{62}
\bibitem[{len()}]{lengthscale}
\bibinfo{note}{As with all `ladder' models, there is an ambiguity in defining
  the value of $l$ because it is also equal to the separation of bases along
  the chain in the double-stranded geometry, whereas in the double helix the
  rise per base pair is 3.3{\AA} due to the wrapping of single strands around
  the axis. As the absolute value of $l$ is used to map onto the experimental
  persistence length of ssDNA we choose to use 6.3{\AA}.}
\end{thebibliography}
\end{document}